\documentclass[runningheads]{llncs}
\usepackage{graphicx}
\usepackage{booktabs}       % professional-quality tables
\usepackage{amsfonts}       % blackboard math symbols
\usepackage{nicefrac}       % compact symbols for 1/2, etc.
\usepackage{microtype}      % microtypography
\usepackage{xcolor}         % colors
\usepackage{mathrsfs}  
\usepackage{bm}
\usepackage{amssymb,amsmath}

\usepackage{subcaption}
\usepackage{wrapfig}
\usepackage{picinpar}
\usepackage{cutwin}
\usepackage{graphicx,multicol}
\usepackage{algorithm}  
\usepackage{algorithmic}
\usepackage{stmaryrd}

\usepackage{multirow}
\usepackage{algorithm}  
\usepackage{algorithmic}
\usepackage{amsfonts}
\usepackage{amsfonts,amssymb} 
\usepackage{verbatim}
\usepackage{url}
\usepackage{enumitem}
\usepackage{hyperref}

\begin{document}
\title{Leapfrog Latent Consistency Model (LLCM) for Medical Images Generation}
\author{Lakshmikar R. Polamreddy \and Kalyan Roy \and Sheng-Han Yueh \and Deepshikha Mahato \and Shilpa Kuppili \and Jialu Li \and Youshan Zhang}

%\author{First Author\inst{1}\orcidID{0000-1111-2222-3333} \and
%Second Author\inst{2,3}\orcidID{1111-2222-3333-4444} \and
%Third Author\inst{3}\orcidID{2222--3333-4444-5555}}
%
\authorrunning{Polamreddy et al.}

\institute{Department of Graduate Computer Science and Engineering, Yeshiva University, 215 Lexington Ave, New York, NY 10016, USA}
%\institute{Princeton University, Princeton NJ 08544, USA \and
%Springer Heidelberg, Tiergartenstr. 17, 69121 Heidelberg, Germany
\maketitle              % typeset the header of the contribution

\begin{abstract}
%%%
The scarcity of accessible medical image data poses a significant obstacle in effectively training deep learning models for medical diagnosis, as hospitals refrain from sharing their data due to privacy concerns. In response, we gathered a diverse dataset named MedImgs, which comprises over 250,127 images spanning 61 disease types and 159 classes of both humans and animals from open-source repositories. We propose a Leapfrog Latent Consistency Model (LLCM) that is distilled from a retrained diffusion model based on the collected MedImgs dataset, which enables our model to generate real-time high-resolution images. We formulate the reverse diffusion process as a probability flow ordinary differential equation (PF-ODE) and solve it in latent space using the Leapfrog algorithm. This formulation enables rapid sampling without necessitating additional iterations. Our model demonstrates state-of-the-art performance in generating medical images. Furthermore, our model can be fine-tuned with any custom medical image datasets, facilitating the generation of a vast array of images. Our experimental results outperform those of existing models on unseen dog cardiac X-ray images. Source code is available at \url{https://github.com/lskdsjy/LeapfrogLCM}.
%%%%
\end{abstract}

%%%% Main Body 
\section{Introduction}

One of the most pressing problems in healthcare is the acute scarcity of comprehensive and diverse datasets for training deep learning models. This issue stems from the hesitance of healthcare institutions to share their data due to privacy concerns and regulatory constraints. As a result, researchers and practitioners face significant challenges in obtaining large-scale, diverse datasets that are essential for training robust deep learning models. This limitation severely hampers the ability to develop models capable of generalizing across various medical conditions and patient demographics, ultimately affecting the quality and effectiveness of AI-driven solutions in healthcare.

The limited availability of comprehensive datasets in the medical field has stymied the progress of AI applications, particularly in training models that can handle the variability and complexity of real-world medical data. The medical domain lacks extensive datasets like ImageNet~\cite{deng2009imagenet} and LAION-5B~\cite{schuhmann2022laion}. This discrepancy has slowed the adoption and effectiveness of AI in medical diagnostics, as models trained on limited data may not perform well across diverse patient populations or rare medical conditions.

In response to this challenge, diffusion models have emerged as a promising solution for generating high-quality and diverse images. Recent advances in diffusion models have demonstrated their ability to produce realistic and varied images, which could potentially address some of the data scarcity issues in healthcare. However, while there has been substantial research into generating high-resolution artistic images, the application of these models to the medical domain remains relatively underexplored. The lack of dedicated datasets and the complexity of medical imaging make it challenging to leverage diffusion models effectively for medical image generation.

\begin{figure*}[t]
\begin{center}
\includegraphics[width=0.9\textwidth]{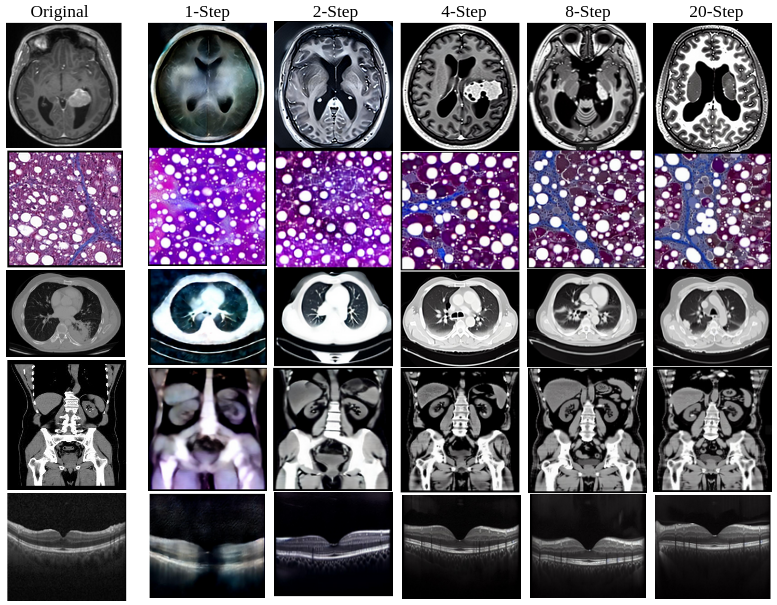}
\caption{Comparison of LLCM different steps generated medical images of (\textbf{512$\times$512}) resolution in different inference steps against original images. } \label{fig:imgs}
\end{center}
\end{figure*}

Our research aims to bridge this gap by introducing a novel contribution to the field. We propose the creation of the \textbf{MedImgs} dataset, which contains a wide range of medical images of both humans and animals. This dataset is designed to provide a comprehensive resource for training and evaluating deep learning models in the medical domain. Additionally, we introduce the leapfrog latent consistency model (\textbf{LLCM}), which utilizes the \textbf{Leapfrog} solver, an efficient and rapid PF-ODE solver, to generate high-resolution images (512$\times$512) in as few as \textbf{1-4} inference steps. These advancements not only address the data scarcity issue but also enhance the performance of deep learning models in medical imaging, paving the way for more accurate and accessible AI-driven medical solutions.

\section{Related Work}\label{sec:rel}
\textbf{Diffusion Models and Samplers.} Diffusion models have been widely applied in image generation tasks. Ho et al.~\cite{ho2020denoising} proposed to synthesize high-quality images using diffusion probabilistic models, a class of latent variable models inspired by considerations from non-equilibrium thermodynamics. Nichol et al.~\cite{nichol2021improved}  proposed a few simple modifications to denoising diffusion probabilistic models (DDPMs)~\cite{ho2020denoising} that can sample much faster and achieve better log-likelihoods with little impact on sample quality. Dhariwal et al.~\cite{dhariwal2021diffusion} proposed to improve sample quality with classifier guidance: a simple, compute-efficient method to trade off diversity for fidelity using gradients from a classifier. Song et al.~\cite{song2020score} proposed to leverage advances in score-based generative modeling to accurately estimate the scores with neural networks, and use numerical SDE (Stochastic Differential Equation) solvers to generate images. Song et al.~\cite{song2020denoising} proposed to accelerate sampling with denoising diffusion implicit models (DDIMs), a more efficient class of iterative implicit probabilistic models with the same training procedure as DDPMs. Rombach et al.~\cite{rombach2022high} proposed latent diffusion models (LDMs) to enable diffusion model (DM) training on limited computational resources while retaining their quality and flexibility. Rombach et al.~\cite{rombach2022text} presented an approach based on retrieval augmented diffusion models (RDMs) to generate high-quality artistic images with text prompts. Song et al.~\cite{song2023consistency} proposed consistency models (CMs), a new family of models that generate high-quality samples by directly mapping noise to data in a single step, and allow multistep sampling to trade compute for sample quality.

\textbf{General Image Synthesis.} There has been significant progress in the field of image synthesis, focusing on controlling various aspects of generated images. Bhunia et al.~\cite{bhunia2023person} proposed to synthesize images of persons based on poses via Denoising Diffusion Model. Lieu et al.~\cite{liu2023more} proposed a unified framework for semantic diffusion guidance to have fine-grained control over image synthesis. Mao et al.~\cite{mao2023guided} proposed initial image editing via diffusion model for guided image synthesis. Wu et al.~\cite{wu2023harnessing} harnessed the spatial-temporal attention of diffusion models for high-fidelity text-to-image (T2I) synthesis. Podell et al.~\cite{podell2023sdxl} proposed SDXL for improving latent diffusion models for high-resolution image synthesis. Xie et al.~\cite{xie2023boxdiff} proposed BoxDiff for T2I synthesis with training-free box-constrained diffusion. Wang et al.~\cite{wang2024compositional} proposed attention mask control strategy based on predicted object boxes for compositional T2I synthesis. Fan et al.~\cite{fan2023frido} proposed Frido, a feature pyramid diffusion model for performing a multi-scale coarse-to-fine denoising process for complex image synthesis. Xue at al.~\cite{xue2023freestyle} leveraged large-scale pre-trained text-to-image diffusion models for layout-to-image synthesis for unseen semantics. Tao et al.~\cite{tao2023galip} proposed generative adversarial CLIPs, namely GALIP, for T2I synthesis by leveraging the powerful pre-trained CLIP model both in the discriminator and generator. Phung et al.~\cite{phung2024grounded} proposed a training-free approach — attention-refocusing — to substantially improve the controllability of T2I synthesis.

\textbf{Medical Image Synthesis.} In the medical domain, image synthesis techniques have evolved through both GAN-based and diffusion-based approaches to generate realistic images for specific medical applications. Uzunova et al.~\cite{uzunova2019multi} proposed a multi-scale patch-based generative adversarial network (GAN) approach to generate large high-resolution 2D and 3D medical images. Armanious et al.~\cite{armanious2020medgan} proposed MedGAN, for medical image-to-image translation which operates on the image level in an end-to-end manner. Havaei et al.~\cite{havaei2021conditional} proposed conditional generative adversarial networks (cGANs) for the conditional generation of medical images via disentangled adversarial inference. Moon et al.~\cite{moon2022multi} proposed Medical Vision Language Learner (MedViLL), which adopts a BERT-based architecture combined with a novel multi-modal attention masking scheme to maximize generalization performance for both vision-language understanding tasks (diagnosis classification, medical image-report retrieval, medical visual question answering) and vision-language generation task (radiology report generation). Pan et al.~\cite{pan20232d} proposed 2D medical image synthesis using transformer-based denoising diffusion probabilistic model. Pinaya et al.~\cite{pinaya2022brain} explored using Latent Diffusion Models to generate synthetic images from high-resolution 3D brain images. Khader et al.~\cite{khader2023denoising} utilized denoising diffusion probabilistic models for 3D medical image generation. Hung et al.~\cite{hung2023med} introduced conditional denoising diffusion probabilistic models (cDDPMs) for medical image generation. Dorjsembe et al.~\cite{dorjsembe2023conditional} proposed conditional diffusion models for semantic 3D medical image synthesis.

While most diffusion models and samplers generate high-resolution images, they typically require 10 or more inference steps to do so. Moreover, diffusion models have predominantly been applied to artistic image synthesis, with relatively few implementations in the medical domain. In the field of medical image synthesis, both GANs and diffusion models have been employed, but their application has been limited to a small number of disease classes, largely due to the lack of large-scale datasets encompassing all human and animal disease categories. To address these limitations, we introduce the Leapfrog Latent Consistency Model (LLCM), a foundation model that is trained on our expansive MedImgs dataset. Our model is capable of generating high-resolution medical images for all disease categories and achieves this with only \textbf{1-4} inference steps, setting a new standard for efficiency and versatility in medical image generation.

\section{Dataset}
We gather medical image data from various online platforms, including Github, Kaggle, and Roboflow Universe, with 78 disease types spanning both human and animal domains, covering diverse body parts. We met several challenges, including varied formats and irrelevant content, as well as few or excess images for a particular category. We removed improper images, corrupted files, and non-image data. We limit each class to 2000 images to prevent bias, with any excess images reserved for testing. Furthermore, 
% we discard classes with fewer than 100 images from training to ensure data integrity. 
we converted all images to JPG format to maintain consistency.
% , while organization and metadata creation facilitated structured dataset preparation. 
After cleaning the data, the disease types were reduced from 78 to 61, and the total distinct classes were 159. Out of 61 disease types, 49 belong to humans, and 12 belong to animals. We name this dataset \textbf{MedImgs}, and it contains a total of 181,117 train images with 159 classes and 69,010 test images with 35 classes. More details of our dataset are shown in Table.~\ref{tab:animal}, and Table.~\ref{tab:human}.

Out of the 61 final disease types, 12 belong to animals, and these cover a wide range of body parts and disease conditions. For example, cattle diseases include conditions affecting the eye (e.g., pink eye and sehat), skin (e.g., lumpy skin disease), and general health (e.g., mouth disease, infected foot). Similarly, other animal categories include chicken diseases such as coccidiosis and salmonella, and fish diseases like columnaris and gill disease. In total, the animal disease category consists of 12 disease types distributed across various species such as cattle, dogs, chickens, and fish. The detailed list of animal diseases can be seen in Table.~\ref{tab:animal}.

The human disease category, which forms the majority of the dataset, covers 49 disease types across multiple body parts. These include conditions like Alzheimer’s disease, brain tumors, breast cancer, lung cancer, and foot ulcers, to name a few. Several specific diseases are well-represented, such as Diabetic Retinopathy, Glaucoma, and Retinal OCT images in the eye category, or various skin conditions such as acne, melanoma, and monkeypox. Additionally, the dataset includes several rare or complex conditions, including Down syndrome, multiple sclerosis, and various forms of cancer (e.g., pancreatic, esophageal, prostate). Table.~\ref{tab:human} provides more detailed information on the disease categories in the human dataset.

After cleaning and preprocessing, the final \textbf{MedImgs} dataset contains a total of 181,117 images for training, covering 159 disease classes. For testing, we set aside 69,010 images across 35 different classes. We maintained a balanced distribution within each class for training to avoid bias toward particular categories. As an example, the Alzheimer's disease class includes images for Mild Demented (896 images), Non-Demented (2000 images), and Very Mild Demented (293 images), while the breast cancer category contains 2000 images each for benign and malignant cases. Other categories, such as bone fracture, have a complete set of 2000 images for both fractured and non-fractured bones.

The test dataset similarly reflects a wide variety of conditions, with several classes including more than 2000 images. For instance, the benign and malignant breast cancer classes in the test set contain 4074 and 4042 images, respectively. Other conditions, like Endometrioid Carcinoma and High-Grade Serous Carcinoma, have larger test sets with 6154 and 11,207 images, respectively. %The supplementary material clearly outlines the number of images in each training and test class. 
This large test set allows for robust model evaluation and ensures that trained models generalize well to unseen data.

\begin{table*}
\begin{center}
\caption{List of the datasets collected for animal disease categories}\label{tab:animal}
\setlength{\tabcolsep}{+0.1cm}{
\begin{tabular}{lcc}
\hline
Animal & Body part & Disease category/Conditions\\
\hline
Cattle & All Parts & \multirow{2}{*}{\begin{tabular}[c]{@{}c@{}}Infected Foot Image, Mouth Disease Infected, Normal Healthy Cow, \\ Normal Mouth Image, Lumpy skin \end{tabular}} \\
\\
Cattle & Eye & Pink eye, Sehat \\
Cattle & Skin & Healthy skin, Lumpy skin \\
Chicken & Egg & Coccidiosis, Salmonella, Newcastle, Healthy \\
Cow & All Parts & FMD, 18K, LSD, NOR \\
Cow & Legs & FMD, Healthy Knuckles, Healthy Legs, Swollen Joints \\
Cow & Skin & FMD, Lumpy Skin, Normal Skin \\
Cow & Teats & Class I, Class 2, Class 3, Class 4 \\
Dog & Eyes & Conjunctival injection/redness, Ocular discharge, Skin lesions \\
Dog & Skin & Keratosis, Nasal discharge, Skin lesions \\
Fish & Skin & Columnaris Disease, EUS Disease, Gill Disease, Healthy Fish, Streptococcus
Disease \\
Hen & Respiratory & Coryza, CRD, Healthy \\
\hline    
\end{tabular}}
\end{center}
\end{table*}

\begin{table*}
\begin{center}
\caption{List of the datasets collected across varied disease categories of human body parts}\label{tab:human}
\setlength{\tabcolsep}{+0.1cm}{
\begin{tabular}{lcc}
\hline
Body part & Disease category\\
\hline
Blood & Acute Lymphoblastic Leukemia(ALL), Malaria \\
Bone & Bone Fracture \\
Brain & Alzheimer's, Cancer, Hemorrhage, Stroke, Tumour, Pediatric Brain Tumour \\ 
Breast & Breast Cancer \\
Cells & Cell Diseases, Lymphoma Cancer \\
Chest & Chest CT-Scan images \\
Digestive System & Gastrointestinal Disease, GERD \\
Esophagus & Esophagitis, Esophageal Cancer \\
Eyes & \multirow{2}{*}{\begin{tabular}[c]{@{}c@{}}Cataract, Diabetic Retinopathy, Eye Diseases, Glaucoma Detection, \\ Retinal OCT Images, Human pink eye virus detection\end{tabular}} \\
\\
Feet/Leg & Foot Ulcer \\
Gene & Down syndrome \\
Hair & Head Lice, Alopacea Areata, Folliculitis, Lichen Planus \\
Heart & Cardiomegaly, Hypertrophic Cardiomyopathy \\
Joints & Arthritis, Osteoarthritis Prediction \\
Kidney & Kidney Stone \\
Knee & Osteoporosis (Knee) \\
Legs & Varicose Detection \\
Liver & Liver Disease \\
Lungs & Lung Cancer, Tuberculosis \\
Mouth & Oral cancer, Oral Diseases \\
Neck & Thyroid \\
Nerve & Multiple Sclerosis \\
Ovaries & PCOS\\
Pancreas & Pancreatic CT Images \\
Prostate & Prostate Cancer \\
Shoulder & Shoulder Implant \\
Skin & \multirow{2}{*}{\begin{tabular}[c]{@{}c@{}}Acne, Atopic Dermatitis, Lyme Disease, Monkey Pox, \\Skin Burn, Skin Diseases, Vitilgo, \\ Tinea Ringworm and other Fungal Diseases, Psoriasis, Melanoma\end{tabular}} \\
\\
\\
Spine & RSNA Cervical Spine \\
Stomach & Gatric Carcinoma(Stomach Cancer) \\
Teeth & Dental Cavity \\
Throat & Pharyngitis Detection \\
Tissue & Gangrene Detection \\
Tongue & Tongue Disease \\

\hline    
\end{tabular}}
\end{center}
\end{table*}

\section{Preliminaries}\label{sec:pre}

\subsection{Forward Process and Diffusion SDEs}
Diffusion models generate data by gradually perturbing data to noise via Gaussian perturbations, then creating samples from noise via sequential denoising steps. Diffusion Probabilistic Models (DPMs)~\cite{ho2020denoising} define a forward process that satisfies: 
$q_{0t}(x_t | x_0) = \mathcal{N}(x_t | \alpha(t) x_0, \sigma^2(t)I)$  where $\alpha(t)$, $\sigma(t) \in \mathbb{R}^+$ are differentiable functions of $t$. Moreover, Kingma et al.~\cite{kingma2021variational} proved that the following stochastic differential equation (SDE) has the same transition distribution $q_{0t}(x_t | x_0)$ for any $t \in [0, T]$:
\begin{equation}\label{eq:dx_t1}
dx_t = f(t) x_t \, dt + g(t) \, dw_t, \quad x_0 \sim q_0(x_0),
\end{equation}
where $w_t \in \mathbb{R}^D$ is the standard Wiener process. Under some regularity conditions, Song et al.~\cite{song2020score} showed that the forward process in Eq.~\ref{eq:dx_t1} has an equivalent reverse process from time $T$ to $0$, starting with the marginal distribution $q_T(x_T)$:

\begin{equation}\label{eq:dx_t_}
dx_t = \left[ f(t) x_t - g^2(t) \nabla_x \log q_t(x_t) \right] dt + g(t)d \overline{w}_t, \quad x_T \sim q_T(x_T),
\end{equation}
where $\overline{w}_t$ is a standard Wiener process in the reverse time. In practice, DPMs use a neural network $\epsilon_\theta(x_t, t)$ parameterized by $\theta$ to estimate the scaled score function: $-\sigma_t\nabla_x \log q_t(x_t)$.

\begin{equation}\label{eq:dx_t}
dx_t = \left[ f(t)x_t + \frac{g^2(t)}{\sigma_t} \epsilon_\theta(x_t, t) \right] dt + g(t)d\overline{w}_t, \quad x_T \sim \mathcal{N}(0, \tilde{\sigma}^2 I).
\end{equation}

 Samples can be generated from DPMs by solving the diffusion SDE in Eq.~\eqref{eq:dx_t} with numerical solvers, which discretize the SDE from $T$ to $0$.

\subsection{Reverse Process and Diffusion (Probability Flow) ODEs}
For faster sampling, we can consider the associated probability flow ODE~\cite{song2020score}, which has the same marginal distribution at each time $t$ as the SDE. Specifically, for DPMs, Song et al. \cite{song2020score} proved that the probability flow ODE of Eq.~\eqref{eq:dx_t} is

\begin{equation}
\frac{dx_t}{dt} = f(t) x_t - \frac{1}{2} g^2(t) \nabla_x \log q_t(x_t), \quad x_T \sim q_T(x_T).
\end{equation}

 By replacing the score function with the noise prediction model, Song et al.~\cite{song2020score} defined the following parameterized ODE (diffusion ODE):

\begin{equation}
\frac{dx_t}{dt} = h_\theta(x_t, t) := f(t)x_t + \frac{g^2(t)}{2\sigma_t}\epsilon_\theta(x_t, t), \quad x_T \sim \mathcal{N}(0, \widetilde{\sigma}^2I).
\end{equation}

 Samples can be drawn by solving the ODE from $T$ to $0$. Compared with SDEs, ODEs can be solved with larger step sizes as they have no randomness.

\subsection{Consistency Models}
The Consistency Model (CM), proposed by Song et al.\cite{song2023consistency}, represents a novel approach for generative models, particularly for one-step or few-step generation tasks. The consistency function \( f \), defined as \( f : (x_t, t) \mapsto x_{\epsilon} \) with \( \epsilon \) denoting a small positive number, is required to satisfy the self-consistency property:
$ f(x_t, t) = f(x_{t'}, t'), \quad \forall t, t' \in [\epsilon, T] $. To facilitate the learning of a consistency model \( f_{\theta} \), the parameterized model enforces the condition \( f_{\theta}(x, \epsilon) = x \). A one-step estimation of \( x_{t_n} \) from \( x_{t_{n+1}} \) is computed using \( \Phi \), where \( \Phi \) represents the one-step PF-ODE solver.

\begin{equation}
    L(\theta, \theta^-; \Phi) = \mathbb{E}_{x,t} \left[ d\left( f_{\theta}(x_{t_{n+1}}, t_{n+1}), f_{\theta^-}(\hat{x}_{\Phi}^{t_n}, t_n) \right) \right], 
\end{equation}
where \( \theta^- \) refers to the target model, \( d(\cdot, \cdot) \) is a chosen metric function for measuring the distance between two samples, e.g., the squared \( \ell_2 \) distance \( d(x, y) = ||x - y||_2^2 \). \( \hat{x}_{\Phi}^{t_n} \) is a one-step estimation of \( x_{t_n} \) from \( x_{t_{n+1}} \) as:
\begin{equation}
    \hat{x}_{\Phi}^{t_n} \leftarrow x_{t_{n+1}} + (t_n - t_{n+1})\Phi(x_{t_{n+1}}, t_{n+1}; \phi). 
\end{equation}

\section{Methods}\label{sec:met}
\begin{figure*}[t]
\begin{center}
\includegraphics[width=1.0\textwidth]{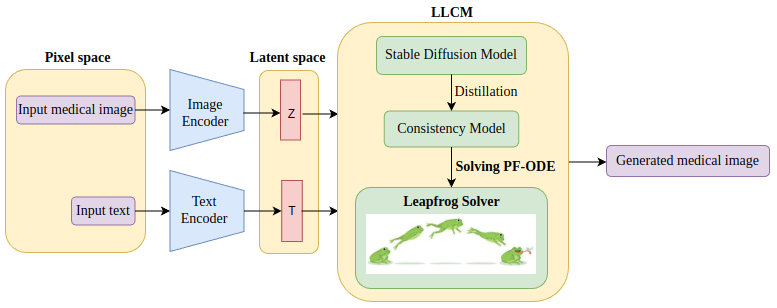}
\caption{Flowchart of Image Generation with Leapfrog Latent Consistency Model.} \label{fig:flowchart}
\end{center}
\end{figure*}

\textbf{Leapfrog Latent Consistency Model.} The flowchart of medical image generation with our model is shown in Fig.~\ref{fig:flowchart}. We use encoders to project an image and its respective text prompt onto latent space (Z, T). Then, we retrain a stable diffusion model with the latent space data. We further distill a consistency model from the retrained stable diffusion model to solve the PF-ODE of the reverse diffusion process with a leapfrog algorithm for generating new images. The PF-ODE of the reverse diffusion process in latent space can be represented in the following equation. 

\begin{equation}\label{eq:dz_t}
\frac{dz_t}{dt} = f(t)z_t + \frac{g^2(t)}{2\sigma_t}\epsilon_\theta(z_t, t), \quad z_T \sim \mathcal{N}(0, \widetilde{\sigma}^2I),
\end{equation}
where $z_t$ are image latents, $\epsilon_\theta (z_t, t)$ is the noise prediction model. As we focus on the conditional generation of images, Eq.~\eqref{eq:dz_t} can be represented as: 
\begin{equation}\label{eq:dz_w}
\frac{dz_t}{dt} = f(t)z_t + \frac{g^2(t)}{2\sigma_t}\epsilon_\theta(z_t, c, t), \quad z_T \sim \mathcal{N}(0, \widetilde{\sigma}^2I),
\end{equation}
where $c$ is the given condition that refers to the text prompt of the image.

 Utilizing Classifier-free guidance(CFG)~\cite{ho2022classifier} is essential for generating high-quality text-aligned images.  Given a CFG scale $\omega$, the original noise prediction is replaced by a linear combination of conditional and unconditional noise prediction, i.e., $\tilde{\epsilon}_{\theta}(z_t, \omega, c, t) = (1 + \omega)\epsilon_{\theta}(z_t, c, t) - \omega\epsilon_{\theta}(z, \phi, t)$.

 If we introduce CFG into the PF-ODE, then Eq.~\eqref{eq:dz_w} becomes:
\begin{equation}
\frac{dz_t}{dt} = f(t)z_t + \frac{g^2(t)}{2\sigma_t}\epsilon_\theta(z_t, \omega, c, t), \quad z_T \sim \mathcal{N}(0, \widetilde{\sigma}^2I).
\end{equation}

 Samples can be generated by solving the PF-ODE from $T$ to $0$. To perform the distillation with a consistency model in latent space, we introduce the consistency function $f_{\theta}: (z_t, \omega, c, t) \mapsto z_0$ to directly predict the solution of PF-ODE for $t = 0$. We parameterize $f_{\theta}$ by the noise prediction model $\hat{\epsilon}_{\theta}$ as:

\begin{equation}
f_{\theta}(z, \omega, c, t) = c_{\text{skip}}(t)z + c_{\text{out}}(t) \cdot \left(\frac{z - \sigma_t \hat{\epsilon}_{\theta}(z, \omega, c, t)} {\alpha_t}\right), \text{ (}\epsilon\text{-Prediction)},
\end{equation}
 where $c_{\text{skip}}(0) = 1$, $c_{\text{out}}(0) = 0$, and $\hat{\epsilon}_{\theta}(z, \omega, c, t)$ is a noise prediction model that initializes with the identical parameters as the retrained diffusion model.

 We utilize the \textbf{Leapfrog} ODE solver $\Psi(z_t, t, s, c)$ for approximating the integration of the right-hand side of Eq.~\eqref{eq:dz_w} from time $t$ to $s$. It is an efficient numerical ODE solver that works by jumping the time steps and making faster convergence.
With this jumping-step technique, our LLCM aims to ensure consistency between the current time step and $k$-step away, $t_{n+k} \rightarrow t_n$. In our main experiments, we set k=20, drastically reducing the length of schedule from thousands to tens.  Our LLCM aims to predict the solution of the PF-ODE by minimizing the consistency distillation loss~\cite{song2023consistency} as given by:

\begin{equation}
\textit{L}_{\textit{{CD}}} (\theta, \hat{\theta}; \Psi) = \mathbb{E}_{z,\omega,c,n} \left[ d  \left( f_{\theta}(z_{t_{n+k}}, \omega, c, t_{n+k}) , f_{\theta^-}(\hat{z}^{{\Psi},{\omega}}_{t_n}, \omega, c, t_n) \right) \right],
\end{equation}
where $\omega$ and $n$ are uniformly sampled from the interval $[\omega_{\text{min}}, \omega_{\text{max}}]$ and $\{1, \ldots, N-1\}$ respectively.
$\hat{z}^{{\Psi},{\omega}}_{t_n}$ is estimated using the new noise model $\tilde{\epsilon}_{\theta}(z_t, \omega, c, t)$, as follows:

\begin{equation}
\hat{z}^{{\Psi},{\omega}}_{t_n} - z_{t_{n+k}} = \int_{t_{n}}^{t_{n+k}} \left( f(t)z_t + \frac{g^2(t)}{2\sigma_t} \tilde{\epsilon}_{\theta}(z_t, \omega, c, t) \right) dt.
\end{equation}

From the above, we get:

\begin{equation}
\hat{z}^{{\Psi},{\omega}}_{t_n} \leftarrow z_{t_n+k} + (1 + \omega) \Psi(z_{t_n+k}, t_{n+k}, t_n, c) - \omega \Psi(z_{t_n+k}, t_{n+k}, t_n, \phi).
\end{equation}

\begin{table*}[t]
\begin{center}
\caption{ Fréchet Inception Distance (FID) scores~\cite{heusel2017gans} for different models across various inference Steps on 35 test classes: This table presents FID values for 175,000 generated images across multiple models and inference steps. Lower FID values indicate higher image quality and better alignment with real data distribution. }\label{tab:fid}
\setlength{\tabcolsep}{+0.2cm}{
\begin{tabular}{lccccccc}
\hline
Model & Step1 & Steps2 & Steps4 & Steps6 & Steps8 & Steps10 & Steps20 \\
\hline
Stable Diffusion~\cite{rombach2022high} & 468.03 & 457.29 & 249.18 & 211.13 & 189.45 & 178.00 & 157.70 \\
%Stable Diffusion (50k)~\cite{rombach2022high} & 466.73	& 456.97 & 245.43 & 206.89 & 185.11 & 173.462 & \\
Dreambooth~\cite{ruiz2023dreambooth} & 488.62 & 466.34 & 300.15 & 250.76 & 220.20 & 205.33 & 186.45 \\
LCM~\cite{luo2023latent} & 256.26 & 246.66 & 243.88 & 240.77 & 238.60 & 237.40 & 237.87 \\
\hline 
\textbf{LLCM(Ours)} & 198.32 & 195.79 & \textbf{145.68} & 168.74 & 191.91 & 198.32 & 185.63 \\
\hline    
\end{tabular}}
\end{center}
\end{table*}

\begin{figure*}[t]
\begin{center}
\includegraphics[width=0.9\textwidth]{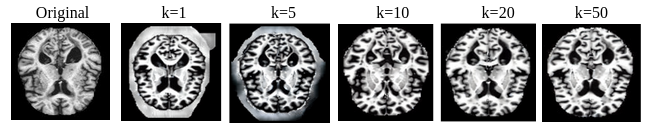}
\caption{Comparison of LLCM generated medical images of (\textbf{512$\times$512}) resolution of the Alzheimer’s mild demented category in humans for different leapfrog jumping steps with 4-step inference.} \label{fig:jumps}
\end{center}
\end{figure*}

\begin{figure*}[h!]
\begin{center}
\includegraphics[width=0.9\textwidth]{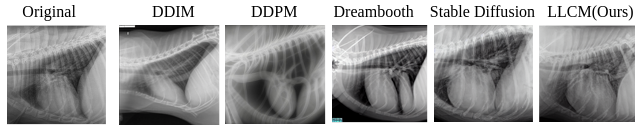}
\caption{Comparison results of unseen large dog heart X-ray images generated by various models.} \label{fig:dog}
\end{center}
\end{figure*}

\begin{figure*}
    \centering
    \includegraphics[width= 0.9\linewidth]{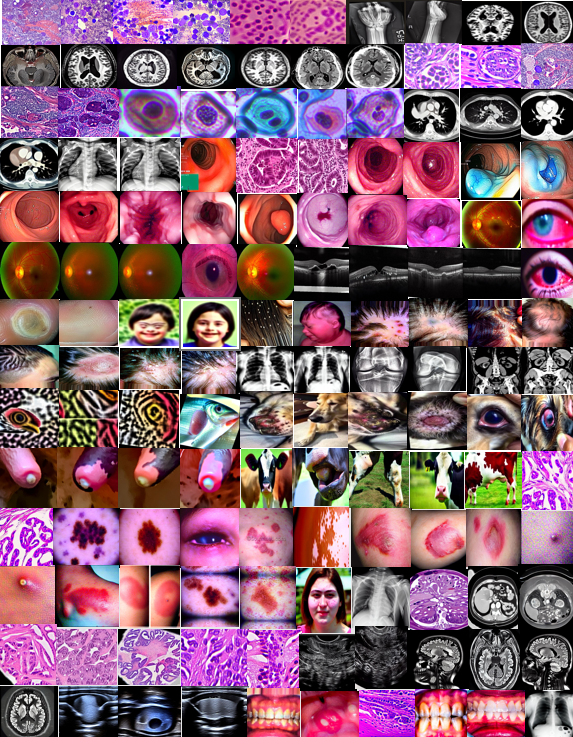}
    \caption{Generated images results with our LLCM model  of (\textbf{512$\times$512}) resolution with 4-step inference.}
    \label{fig:more_imgs}
\end{figure*}

We solve the above equation with the Leapfrog Solver.
Given the position of a particle at $x_1$ and the velocity at the next midpoint $ v_{3/2}$ are determined by the equations:
\begin{equation}
    x_1 = x_0 + hv_{1/2}, \quad v_{3/2} = v_{1/2} + hF(x_1),
\end{equation}
 where \( x_1 \) is the position in the next time step,
\( x_0 \) is the initial position,
\( h \) is the interval between two time steps,
\( v_{1/2} \) is the velocity at the midpoint, and 
$F(x)= \frac{dv}{dt} $. Then we can step forward to $x$ with  $x_2 = x_1 + hv_{3/2}$. We solve the reverse PF-ODE with the Leapfrog approach by approximating the initial position ($x_0$) and initial velocity ($v_0$) terms based on the DDIM paper~\cite{song2020denoising},
$x_t = \sqrt{\alpha_{t-1}} \cdot \hat{x}_0$, $v_t = \sqrt{1 - \alpha_{t-1}} \cdot \hat{\epsilon}$, and $v_{1/2} = 2v_t$:
\begin{align}
\hat{x}_{t-1} &= x_t + hv_{1/2},
\end{align}
 where $\hat{x}_0$, $\hat{\epsilon}$ are predicted by the model, $x_{t}$ is the noised image at the time step t, and $\hat{x}_{t-1}$ is the image at the previous time step approximated by the solver in a single iteration.

 The pseudo-code for the training of our LLCM model is provided in Alg.~\ref{alg:llcm}. We randomly select two-time steps $t_n$ and $t_{n+k}$ that are $k$ time steps apart and apply the same Gaussian noise $\epsilon$ to obtain the noised data $z_{tn}$, $z_{tn+k}$ as 
$z_{tn+k} \leftarrow \alpha(t_{n+k})z + \sigma(t_{n+k})\epsilon$, $z_{tn} \leftarrow \alpha(t_{n})z + \sigma(t_{n})\epsilon$.
Then, we can directly calculate the consistency loss for these two-time steps. Notably, this method can also utilize the jumping-step technique to speed up the convergence.

\begin{algorithm}
\textbf{Input: }{Dataset $D(s)$, re-trained LLCM parameter $\theta$, learning rate $\eta$, distance metric $d(\cdot, \cdot)$, EMA rate $\mu$, noise schedule $\alpha(t)$, $\sigma(t)$, guidance scale $[w_{\text{min}}, w_{\text{max}}]$, jumping interval $k$, and encoder $E(\cdot)$}
Encode training data into the latent space: $D(s)$ \\
$z = \{(z, c) | z = E(x), (x, c) \in D(s)\}$ \\
$\theta^- \leftarrow \theta$ \\
\textbf{Repeat} \\
\hspace*{1em}Sample $(z, c) \sim D(s)$ \\
\hspace*{1em}$z, n \sim U[1, N - k]$ and $w \sim [w_{\text{min}}, w_{\text{max}}]$ \\
\hspace*{1em}Sample $\epsilon \sim N(0, I)$ \\
\hspace*{1em}$z_{tn+k} \leftarrow \alpha(t_{n+k})z + \sigma(t_{n+k})\epsilon$, $z_{tn} \leftarrow \alpha(t_{n})z + \sigma(t_{n})\epsilon$ \\
\hspace*{1em}$L(\theta, \theta^-) \leftarrow d(f_{\theta}(z_{tn+k}, t_{n+k}, c, w), f_{\theta^-}(z_{tn}, t_{n}, c, w))$ \\
$\hspace*{1em}\theta \leftarrow \theta - \eta\nabla_{\theta} L(\theta, \theta^-)$ \\
\hspace*{1em}$\theta^- \leftarrow \text{stopgrad}(\mu\theta^- + (1 - \mu)\theta)$ \\
\textbf{Until Convergence}
\caption{Leapfrog Latent Consistency Model (LLCM)}\label{alg:llcm}
\end{algorithm}

\section{Experiments and Results}\label{sec:exp}

\textbf{Implementation details.}
Our model has been trained on 
\textbf{eight} NVIDIA A100-SXM4-80GB GPUs, renowned for their unparalleled performance in deep learning tasks, with the MedImgs dataset employing crucial hyper-parameters to ensure optimal performance. Specifically, the training spans across 55 epochs, with each epoch comprising 184 batches. Within each batch, 128 samples per device are processed, resulting in a substantial total train batch size of 1024. These efforts are supported by the Adam optimizer, leveraging a learning rate of 8e-6. Additionally, an exponential moving average decay of 0.95 is employed to stabilize training and enhance model convergence. Gradient accumulation steps are set to 1, and the entire training process encompasses 10,000 iterations. It takes only 24 hours with eight NVIDIA A100 GPUs.

% \subsection{Results}\label{sec:res}
\textbf{Results.} 
To evaluate the performance of our proposed Leapfrog Latent Consistency Model (LLCM), we generate 5,000 images of (\textbf{512$\times$512}) resolution for each of the 35 test classes, resulting in a total of 175,000 generated images per inference step. The inference steps range from step 1, step 2, step 4, step 6, step 8, step 10, to step 20. For comparison, we also generate 175,000 images using several other diffusion models, including Stable Diffusion~\cite{rombach2022high}, Dreambooth~\cite{ruiz2023dreambooth}, and Latent Consistency Model (LCM)~\cite{luo2023latent}. To quantify the quality of the generated images and their alignment with the real data distribution, we employ the widely-used Fréchet Inception Distance (FID) metric~\cite{heusel2017gans}. The Table.~\ref{tab:fid} illustrates FID comparisons of these models across different inference steps. The LLCM model achieves superior performance, particularly at earlier inference steps. At Step 4, LLCM achieves the lowest FID score of 145.68, which demonstrates its ability to generate high-quality images with fewer inference steps, significantly outperforming both Stable Diffusion and Dreambooth. In comparison, the FID score of Stable Diffusion at Step 4 is 249.18, while Dreambooth shows a much higher FID of 300.15, indicating that LLCM generates images that are closer to the real data distribution with enhanced consistency and quality. LCM, though efficient, shows minimal improvement in FID values as the number of steps increases. On the other hand, LLCM shows a notable drop in FID scores at Step 4, followed by a moderate increase at later steps, indicating that LLCM performs best with fewer steps, aligning well with scenarios requiring faster inference without compromising image quality. The results suggest that LLCM's leapfrog mechanism allows for improved synthesis of images in the early inference steps compared to other models, highlighting its potential for efficient image generation in scenarios where computational resources or time are constrained.

Furthermore, we conduct an ablation study with various jumping steps of the Leapfrog solver to determine the optimal jumping step value $ k$. Fig.~\ref{fig:jumps} reveals that we can get the optimal image with $k = 20 $ for the Alzheimer’s mild demented category in humans across different jumping steps with 4-step inference. To test the generalization ability of our model, we applied our LLCM model in an unseen dog heart X-ray dataset~\cite{li2024regressive}, and we achieved remarkable results compared to state-of-the-art generation models. Figure~\ref{fig:dog} illustrates a comparison between images generated by various models and the original image.

In addition, Fig.~\ref{fig:more_imgs} displays several images generated by our LLCM model using a 4-step inference process, covering various human and animal diseases. Notably, we observe that the image quality for human diseases is consistently higher compared to that of animal diseases. This can be attributed to the larger availability of training data for human diseases, leading to better model generalization and more refined image synthesis in these categories. This discrepancy in performance can be attributed to the relatively smaller number of training images available for animals.

\section{Conclusion}\label{sec:con}
We address the challenge of data scarcity in medical diagnosis by introducing MedImgs, a diverse dataset featuring both human and animal medical images. Furthermore, our LLCM model demonstrates superior performance in generating high-quality images with minimal inference steps compared to existing state-of-the-art models. Our future efforts will focus on expanding the dataset to include more medical images and enhancing the model to produce high-quality images in a single inference step.

\bibliographystyle{splncs04}
\bibliography{mybibliography}




% LNCS Homepage, \url{http://www.springer.com/lncs}. Last accessed 4
% \end{thebibliography}
\end{document}